\documentclass[aps,prd,onecolumn,groupedaddress,showpacs,nofootinbib,amssymb]{revtex4}
\usepackage[dvips]{graphicx}
\usepackage{amssymb}
\usepackage{amsmath}
\usepackage{graphicx,,color}
\usepackage{amsfonts}
\usepackage{bm}
\usepackage{cancel}
\usepackage{comment}
\newcommand\be{\begin{equation}}
\newcommand\ee{\end{equation}}

\allowdisplaybreaks[4]

\begin{document}

\title{Unification of Inflation with Dark Energy in $f(R)$ Gravity and Axion Dark Matter}
%\title{Inflation and Eschatological Scenarios with $f(R)$ Gravity and Axion Dark Matter}
\author{S.D. Odintsov,$^{1,2,3}$\,\thanks{odintsov@ieec.uab.es}
V.K. Oikonomou,$^{4,5,3}$\,\thanks{v.k.oikonomou1979@gmail.com}}
\affiliation{$^{1)}$ ICREA, Passeig Luis Companys, 23, 08010 Barcelona, Spain\\
$^{2)}$ Institute of Space Sciences (IEEC-CSIC) C. Can Magrans s/n,
08193 Barcelona, Spain\\
$^{3)}$ Tomsk State Pedagogical University, 634061 Tomsk, Russia\\
$^{4)}$Department of Physics, Aristotle University of Thessaloniki, Thessaloniki 54124, Greece\\
$^{5)}$ Laboratory for Theoretical Cosmology, Tomsk State University
of Control Systems
and Radioelectronics, 634050 Tomsk, Russia (TUSUR)\\
}

\tolerance=5000

\begin{abstract}
In this work we introduce an effective model of $f(R)$ gravity
containing a non-minimal coupling to the axion scalar field. The
axion field is described by the misalignment model, in which the
primordial $U(1)$ Peccei-Quinn symmetry is broken during inflation
and the $f(R)$ gravity is described by the $R^2$ model, and in
addition, the non-minimal coupling has the form $\sim
h(\phi)R^{\gamma}$, with $0<\gamma<0.75$. By appropriately
constraining the non-minimal coupling at early times, the axion
field remains frozen in its primordial vacuum expectation value,
and the $R^2$ gravity dominates the inflationary era. As the
Universe expands, when $H$ equals the axion mass $m_a$ and for
cosmic times for which $m_a\gg H$, the axion field oscillates. By
assuming a slowly varying evolution of the axion field, the axion
energy density scales as $\rho_a\sim a^{-3}$, where $a$ is the
scale factor, regardless of the background Hubble rate, thus
behaving as cold dark matter. At late times, the axion still
evolves as $\rho_a\sim a^{-3}$, however the Hubble rate of the
expansion and thus the dynamical evolution of the Universe is
controlled by terms containing the higher derivatives of $\sim
R^{\gamma}$, which are related to the non-minimal coupling, and as
we demonstrate, the resulting solution of the Friedman equation at
late times is an approximate de Sitter evolution. The late-time de
Sitter Hubble rate scales as $H\sim \Lambda^{1/2}$, where
$\Lambda$ is an integration constant of the theory, which has its
allowed values very close to the current value of the cosmological
constant. Finally, the theory has a prediction for the existence
of a pre-inflationary primordial stiff era, in which the energy
density of the axion scales as $\rho_a\sim a^{-6}$.
\end{abstract}

%PACS numbers: 04.50.Kd, 95.36.+x, 98.80.-k, 98.80.Cq

\maketitle

\section{Introduction}

For decades fundamental scalar fields were an important element of
quantum field theory, however these remained undetected until the
Higgs particle was discovered in 2012 \cite{Aad:2012tfa}. This
experimental verification of the theory showed that the spin zero
elementary particles play some important role in model building,
and it is possible that scalar fields may manifest themselves in
other physical phenomena and processes. In cosmology, scalar
fields have a prominent role, since quite many inflationary
theories are based on the slow-rolling of the inflaton field
\cite{Guth:1980zm,Linde:1983gd}. On the other hand alternative
theoretical models of modified gravity can also describe the
inflationary era, see for example Ref. \cite{Starobinsky:1982ee}
for the pioneer Starobinsky model. Modified gravity theory is also
capable to describe dark energy
\cite{Capozziello:2002rd,Carroll:2003wy} and in addition, it can
also successfully provide a unified description of early-time
acceleration with late-time acceleration, see for example
\cite{Nojiri:2003ft}, for the first proposal towards this
direction in the context of $f(R)$ gravity. Several viable $f(R)$
gravities unifying inflation with dark energy were also developed
in Refs. \cite{Nojiri:2007as,Nojiri:2007cq,Cognola:2007zu}, and
also the possibility of incorporating the radiation and matter
domination eras, was addressed in Ref. \cite{Nojiri:2006gh}. For
some recent updates in the field of modified gravity, the reader
is referred to the reviews
\cite{reviews4,reviews5,reviews6,reviews1,reviews2,reviews3}.
Nevertheless, the nature of dark energy and especially of dark
matter as well as their interaction still remain unclear. Dark
matter has been believed for many years that consists of weakly
interacting massive particles, and there are many observational
motivations towards believing in the particle nature of dark
matter, like the collision of galaxies observed in the Bullet
Cluster. Up to date many theoretical proposals were introduced
that could potentially make the observation of dark matter
particles possible, having a wide range of masses, see for example
Ref. \cite{Oikonomou:2006mh} which focuses on the direct detection
of dark matter candidates. However, many experiments up to date
have not achieved in detecting any dark matter particle. Axion
physics is a timely theoretical framework
\cite{Marsh:2015xka,Marsh:2017yvc,Odintsov:2019mlf,Cicoli:2019ulk,Fukunaga:2019unq,Caputo:2019joi,maxim},
due to the related experiments which have gained a lot of
attention and many of the well motivated proposed experiments may
reveal whether the axion exists
\cite{Du:2018uak,Henning:2018ogd,Ouellet:2018beu,Safdi:2018oeu,Rozner:2019gba,Avignone:2018zpw,Caputo:2018vmy,Caputo:2018ljp},
see also \cite{Lawson:2019brd} for an insightful approach. Some of
the proposals indicate that the axion mass could be extremely low,
of the order $m_a\sim \mathcal{O}(10^{-12})$eV, and it is well
known from the axion physics literature that axions with such a
low mass could be perfect candidates for dark matter, at least
some of it in the Universe \cite{Marsh:2015xka}. In fact, the most
promising model for axion is the so-called misalignment axion
model, which is based on a broken Peccei-Quinn $U(1)$ symmetry
during inflation \cite{Dine:2004cq}, in contrast to the QCD axion
models which have unbroken Peccei-Quinn $U(1)$ symmetry during
inflation. Most of the misalignment models can be string theory
motivated, since several string theory models predict such
low-mass axion particles \cite{Marsh:2015xka}. The detection of
the axion particle is quite easy to realize and it is based on the
fact that the axion can interact with photons in the presence of
magnetic fields
\cite{Balakin:2009rg,Balakin:2012up,Balakin:2014oya}, thus leaving
many possibilities of detection. To our opinion, the last
forefront of dark matter, before the whole context collapses, is
low mass particles like axions or even neutrinos, unless
supersymmetry is confirmed experimentally by detecting some
supersymmetric partner in the Large Hadron Collider.

It is interesting to note that among different dark matter models,
one can mention also an $f(R)$ gravity related dark matter model
presented in Ref. \cite{Capozziello:2006ph}. Furthermore, a
theoretical attempt to unify inflation with dark energy and dark
matter, using scalar Lagrange multipliers, can be found in Ref.
\cite{Nojiri:2016vhu}.

In view of the dark matter and dark energy problem, in this paper
we shall introduce an effective $f(R)$ gravity theory in the
presence of a misalignment axion scalar field. The theory we shall
present has a non-minimal coupling between the axion scalar field
and a curvature dependent function $G(R)$, of primordial origin.
The exact form of the $f(R)$ gravity will contain the $R^2$ model
and this extra non-minimal coupling, with the function $G(R)$
being chosen as $G(R)\sim R^{\gamma}$, $0<\gamma<0.75$. The
misalignment axion model predicts that the axion is frozen at
early times to its vacuum expectation value obtained by the
breaking of the primordial $U(1)$ Peccei-Quinn symmetry, thus it
is dominated dynamically by the $f(R)$ gravity at early times.
Thus during inflation, the leading order terms of the $f(R)$
gravity dominate the evolution, and drive inflation. As the
Universe expands, when $m_a\sim H$, with $H$ being the Hubble
rate, the axion starts to oscillate and when $m_a\gg H$, the axion
oscillates in a slow-varying way. Focusing on later and late
times, where the curvature of the Universe is too low, the
slow-varying assumption leads to the fact that its energy density
scales as $\rho_a\sim a^{-3}$, where $a$ is the scale factor,
regardless of the cosmological background in terms of the Hubble
rate. This is a crucial observation, and as we demonstrate at late
times, the background cosmological evolution in terms of the
Hubble rate, is controlled by the non-minimally coupling term
containing $G(R)\sim R^{\gamma}$. By solving the resulting
Friedman equation at leading order, we show that a de Sitter
expansion is achieved at late times of the form,
\begin{equation}\label{latetimedesitterevolutinaxionfirsttime}
H(t_0)\simeq \frac{\sqrt{2} \sqrt{1-\gamma } \sqrt{\Lambda
}}{\sqrt{3-4 \gamma }}\, ,
\end{equation}
which clearly describes and accelerating expansion, thus the model
successfully generates a dark energy era. Intriguingly enough, the
integration constant $\Lambda$ which emerges from the theory has
dimensions eV$^2$ and it scales as $\Lambda \sim H_0^2$. In
addition, for the allowed values of the parameter $\gamma$, which
are $0<\gamma <0.75$, and particularly when $0<\gamma\leq 0.74$,
and for $H_0$ taking the current observed value of the Hubble rate
$H_0\sim 10^{-33}$eV, the allowed values of the integration
constant $\Lambda$ are $1.5\times 10^{-66}\mathrm{eV}^2<\Lambda <
7.69231\times 10^{-68}$eV$^2$, which are very close to the current
allowed value of the cosmological constant  $\Lambda_0\sim
10^{-66}$eV$^2$. However, as $\gamma$ approaches the value $0.75$,
then the constant $\Lambda$ takes smaller values, for example when
$\gamma=0.74999999999$, then $\Lambda=8\times 10^{-77}$eV$^2$.

Thus with the present effective model we aim to demonstrate that
an $f(R)$ gravity in the presence of a misalignment axion field,
can describe the inflationary era, a dark matter component which
scales at intermediate and late times as $a^{-3}$, and can also
generate the late-time acceleration in the Universe. In addition,
in our model, the primordial non-minimal axion coupling to the
curvature can affect drastically only the late-time dynamics,
since in all previous eras it had no particular effect. Finally,
we shall show that the present model predicts a stiff matter era
preexisting the inflationary era.

In all the following considerations, the background metric will
assumed to be that of a flat Friedmann-Robertson-Walker (FRW),
with line element,
\begin{equation}
\label{metricfrw} ds^2 = - dt^2 + a(t)^2 \sum_{i=1,2,3}
\left(dx^i\right)^2\, ,
\end{equation}
where $a(t)$ is as usual the scale factor.

\section{$f(R)$ Gravity with Axion Dark Matter}

In this work we shall consider a vacuum $f(R)$ gravity theory in
the presence of an axion dark matter scalar field, which is a
canonical scalar field, and we shall assume a non-minimal coupling
between the scalar field and higher powers of the Ricci scalar.
The gravitational action is the following,
\begin{equation}
\label{mainaction} \mathcal{S}=\int d^4x\sqrt{-g}\left[
\frac{1}{2\kappa^2}f(R)+\frac{1}{2\kappa^2}h(\phi)G(R)-\frac{1}{2}\partial^{\mu}\phi\partial_{\mu}\phi-V(\phi)
\right]\, ,
\end{equation}
where $\kappa^2=\frac{1}{8\pi G}$, with $G$ Newton's gravitational
constant. For notational simplicity in deriving the gravitational
equations of motion, we set,
\begin{equation}\label{gravisetnotation}
\mathcal{F}(R,\phi)=\frac{1}{\kappa^2}f(R)+\frac{1}{\kappa^2}h(\phi)G(R)\,
.
\end{equation}
We shall use the metric formalism, so upon variation of the action
(\ref{mainaction}) with respect to the metric tensor, for the FRW
metric of Eq. (\ref{metricfrw}), we get the following
gravitational equations,
\begin{align}\label{eqnsofmkotion}
& 3 H^2F=\frac{1}{2}\dot{\phi}^2+\frac{RF-\mathcal{F}+2
V}{2}-3H\dot{F}\, ,\\ \notag &
-3FH^2-2\dot{H}F=\frac{1}{2}\dot{\phi}^2-\frac{RF-\mathcal{F}+2
V}{2}+\ddot{F}+2H\dot{F}\, ,
\end{align}
where $F=\frac{\partial \mathcal{F}}{\partial R}$. Upon varying
the gravitational action (\ref{mainaction}) with respect to the
scalar field, we obtain the following equation of motion for the
scalar field,
\begin{equation}\label{scalarfieldeqn}
\ddot{\phi}+3H\dot{\phi}+\frac{1}{2}\left(-\mathcal{F}'(R,\phi)+2V'(\phi)\right)=0\,
,
\end{equation}
where the ``prime'' denotes differentiation with respect to the
scalar field $\phi$.

The choice of the $f(R)$ gravity is the following,
\begin{equation}\label{starobinsky}
f(R)=R+\frac{1}{36H_i^2}R^2\, ,
\end{equation}
so it is basically chosen to be the well-known $R^2$ model of
inflation, with the $H_i$ units being eV. As for the non-minimally
coupled term $\sim h(\phi)G(R)$, the exact choice will depend on
the phenomenology of the axion field and on the late-time
behavior. The $h(\phi)$ will be assumed to be inverse proportional
to the scalar field,
\begin{equation}\label{hphichoice}
h(\phi)\sim\frac{1}{\phi^{\delta}}\, ,
\end{equation}
with $\delta>0$, or similar and the reasoning for this choice will
be explained in the next section. Also the $G(R)$ function will be
assumed to have the following form,
\begin{equation}\label{GRfunction}
G(R)\sim R^{\gamma}\, ,
\end{equation}
where $\gamma$ takes values in the interval $0<\gamma <0.75$.

%%%%%%%%%%%%%%%%%%%%%%%%%%%%%%%%%%%%%%%%%%%%%%%%%%%%%%%%%%%%%%%%%%%%%%%%%%%%%%%%%%%%%%%%%%%%%%%%%%%%%%%%%%%%%%%%%%%%%%%%%%%%%%%%%%%

\section{Unified Description of Inflation and Dark Energy}

\subsection{The Evolution of the Axion Field}

The evolution of the axion dark matter scalar field is essential
for the theory, and everything is based on two intrinsic
mechanisms of the axion scalar field, the primordial broken $U(1)$
Peccei-Quinn like symmetry, and the axion field slow-varying
reheating. Basically, the broken Peccei-Quinn symmetry refers to
an era where non-perturbative sting theory phenomena take place
and control the dynamics of the axion. A recent informative review
on the axion field dynamics in the context of various theories can
be found in Ref. \cite{Marsh:2015xka}. We shall adopt many of the
conventions and notation of Ref. \cite{Marsh:2015xka}. One of the
most sound phenomenological models for the axion field, that can
mimic dark matter is the misalignment model, which eventually can
describe some of the dark matter present in our Universe at
present. One additional assumption crucial for the dynamical
evolution of the theory we develop in this paper, is that the term
containing the non-minimal coupling of the axion to the curvature
scalar, namely $\sim h(\phi)G(R)$, satisfies the following
condition during the inflationary era and for all the cosmological
eras that follow,
\begin{equation}\label{conditionforaxion}
2V'(\phi)\gg \mathcal{F}'(R,\phi)\, ,
\end{equation}
or equivalently,
\begin{equation}\label{conditionforaxion1}
2V'(\phi)\gg \frac{1}{\kappa^2}h'(\phi)G(R)\, .
\end{equation}
By integrating Eq. (\ref{conditionforaxion1}) with respect to the
scalar field $\phi$, we obtain,
\begin{equation}\label{conditiononaxion2}
2V(\phi)\gg \frac{1}{\kappa^2}h(\phi)G(R)\, ,
\end{equation}
which is very important for the considerations that will follow on
the dynamical evolution of the theory. The condition
(\ref{conditionforaxion}) is the only assumption of the theory on
the dynamics of the scalar, and clearly restricts severely the
functional form of the non-minimal coupling $h(\phi)$. However,
the behavior of $h(\phi)$ and $G(R)$ made in Eqs.
(\ref{hphichoice}) and (\ref{GRfunction}) can satisfy the
constraint (\ref{conditionforaxion}). Essentially we need a small
contribution of the term $\sim h(\phi)G(R)$ during and after the
inflationary era. However there will be a pre-inflationary era for
which $2V(\phi)\sim  \frac{1}{\kappa^2}h(\phi)G(R)$. During this
pre-inflationary era, the axion field will not behave as dark
matter, and we discuss this issue more concretely in a later
subsection.

In view of the condition (\ref{conditionforaxion}), the equation
of motion for the axion scalar field (\ref{scalarfieldeqn}) is
mainly affected by the axion scalar potential, during and after
the inflationary era. Thus the phenomenology of the misalignment
model applies, in the context of which the primordial $U(1)$
Peccei-Quinn  symmetry is broken during the pre-inflationary era
and during all the subsequent cosmological eras, leaving the axion
field having a vacuum expectation value during the inflationary
era. During the inflationary era, the axion mass $m_a$ is
constant, and this holds true for all cosmic times for which $H\gg
m_a$, where $H$ is the Hubble rate. The axion field potential has
the following form,
\begin{equation}\label{axionpotential}
V(\phi(t))\simeq \frac{1}{2}m_a^2\phi^2_i(t)\, ,
\end{equation}
where $\phi_i$ is the value of the axion during the corresponding
cosmic era. The axion field is overdamped during inflation, and
the following initial conditions hold true for the values of the
axion and its derivatives during inflation,
\begin{equation}\label{axioninitialconditions}
\dot{\phi}(t_i)=\zeta \ll 1,\,\,\,\phi(t_i)=f_a\theta_a\, ,
\end{equation}
where $t_i$ is the cosmic time during the inflationary period,
$f_a$ stands for the axion decay constant and $\theta_a$ is the
so-called initial misalignment angle. Due to the initial
conditions, the axion is frozen during inflation and essentially
contributes a small cosmological constant term during inflation,
as we also show shortly. Basically, the effective equation of
state (EoS) parameter for the axion $w_a$ is approximately
$w_a\sim -1$, so it contributes an effective cosmological term,
and this is due to the fact that it is overdamped. As we show
shortly in the following subsection, the cosmological dynamics or
equivalently the Hubble rate $H(t)$, will be strictly determined
by the $f(R)$ gravity during inflation. Due to the overdamping at
early-times during inflation, we have $\dot{\phi}=\zeta\ll 1$ and
also $\ddot{\phi}=\lambda \ll 1$, and this behavior continues
until $H\sim m_a$, at which point the axion starts to oscillate.
Before we continue with the description of the axion dynamics
during the era $H\sim m_a$, let us first fix the values of the
quantities that appear in the above equations. Firstly, we shall
assume that the inflationary scale will be
$H_I=\mathcal{O}(10^{13})$GeV, so the low-scale inflationary
scenario is realized. Also for phenomenological reasons, the most
plausible values for $f_a$, $\theta_a$ and $m_a$ are
\cite{Marsh:2015xka},
\begin{equation}\label{axionconstraint}
f_a\sim \mathcal{O}(10^{11})\mathrm{GeV},\,\,\, \theta_a \sim
\mathcal{O}(1)\, ,
\end{equation}
\begin{equation}\label{massaxion}
m_a\sim \mathcal{O}(10^{-12})\mathrm{eV}\, .
\end{equation}
In order for the condition (\ref{conditiononaxion2}) to hold true
during the inflationary era, the coupling function $h(\phi)$ and
the function $G(R)$ must have appropriate forms, so let us have an
idea on how these behave during inflation for the above choices of
the parameters, and compare these to the scalar potential. For the
axion field parameter values (\ref{axionconstraint}) and
(\ref{massaxion}), the potential term $2 \kappa^2 V(\phi_i)$
during inflation is of the order $\mathcal{O}(3\times
10^{-38})$eV, while for $\delta=\mathcal{O}(3)$ and for
$0<\gamma<0.75$, the term $\sim h(\phi)G(R)$ during inflation, is
of the order $\mathcal{O}(2\times 10^{-88})$eV, so for this simple
qualitative choice, the constraint (\ref{conditiononaxion2}) holds
true during inflation.

As the Universe evolves and the Hubble rate values drop, after
$m_a\sim H$, and for cosmic times for which $m_a\gg H$, the axion
field has already started oscillating. Actually the oscillations
start when $m_a\sim H$, and this can be viewed as an inherent
reheating mechanism in the theory. In view of the assumption
(\ref{conditionforaxion}), the scalar field equation of motion
when becomes approximately,
\begin{equation}\label{scalarfieldeqnduringradiation}
\ddot{\phi}+3H\dot{\phi}+m_a^2\phi=0\, .
\end{equation}
One can seek slowly varying oscillating solutions of the form
\cite{Marsh:2015xka},
\begin{equation}\label{solutionaxionradandaft}
\phi (t)= A(t)\cos (m_a t)\, ,
\end{equation}
where $A(t)$ is a slowly-varying function of the cosmic time,
which satisfies the constraints,
\begin{equation}\label{Atsolutionconstraints}
\frac{\dot{A}}{m_a}\sim \frac{H}{m_a}\sim \epsilon\ll 1\, .
\end{equation}
Hence, plugging Eq. (\ref{solutionaxionradandaft}) in Eq.
(\ref{scalarfieldeqnduringradiation}), and working at leading
order in $\epsilon$ we obtain the following equation,
\begin{equation}\label{scalarfieldeqnduringradiation1}
-\frac{2 \dot{A}(t) \sin (m_a t)}{m_a}-\frac{3 A(t) H(t) \sin (m_a
t)}{m_a}=0\, ,
\end{equation}
so by expressing $H=\frac{\dot{a}}{a}$, we get,
\begin{equation}\label{newconditionforA}
\frac{d A}{A}=-\frac{3d a}{2a}\, ,
\end{equation}
which when solved yields,
\begin{equation}\label{solutionforA}
A\sim a^{-3/2}\, .
\end{equation}
The axion energy density is,
\begin{equation}\label{axionenergydensity}
\rho_a\sim \frac{1}{2}\dot{\phi}^2+\frac{1}{2}m_a^2\phi^2\, ,
\end{equation}
where we omitted the non-minimal coupling contribution due to the
condition (\ref{conditiononaxion2}). So in view of the solution
(\ref{solutionaxionradandaft}) we get,
\begin{equation}\label{solutionforenergy1}
\rho_a=m_a^2 A(t)^2+\frac{1}{2} dot{A}(t)^2 \cos ^2(m_a t)-m_a
A(t) \dot{A}(t) \sin (m_a t) \cos (m_a t)\, ,
\end{equation}
so at leading order we have \cite{Marsh:2015xka},
\begin{equation}\label{leadingorderrhoa}
\rho_a\sim A^2\, ,
\end{equation}
or in view of the solution (\ref{solutionforA}) we have
equivalently,
\begin{equation}\label{leadingorderrhoa1}
\rho_a\sim a^{-3}\, .
\end{equation}
The above result is of crucial importance since it shows that the
axion field dynamics after the frozen state during inflation, is
that of cold dark matter, regardless of the Hubble rate
background. Thus the actual cosmological evolution $H(t)$ is not
affected by the scalar field, and in our model the $f(R)$ gravity
will determine the inflationary and late-time dynamics of the
spacetime, with the axion field evolving as dark matter. Actually,
for all the subsequent eras after inflation, for which $H\ll m_a$,
which corresponds to cosmic times $t\gg 1/m_a$, the averaged EoS
parameter for the axion scalar field is $\langle
w_{eff}\rangle\sim 0$ due to the sinusoidal time dependence
\cite{Marsh:2015xka}, and this holds true regardless of the
background evolution, thus for any Hubble rate. This is important
and this result holds true for any cosmic time for which $m_a\gg
H$, so well after the inflationary era.

In conclusion, the axion field during the inflationary era is
frozen to its string originating vacuum expectation value acquired
by the primordial pre-inflationary breaking of the Peccei-Quinn
symmetry, and well after inflation it oscillates in a way so that
its energy density evolves as $\rho_a\sim a^{-3}$, so it evolves
as cold dark matter, regardless the background evolution. Thus the
background evolution is controlled by the $f(R)$ gravity, both
during inflation and at late-times, as we demonstrate in a
quantitative way in the following two subsections.

\subsection{Background Evolution During the Inflationary Era: The Role of the $f(R)$ Gravity}

In this section we shall discuss the essential features of the
inflationary era for the $f(R)$ axion model. As we already
mentioned in the previous subsection, the axion field is frozen
during the inflationary era, so it does not affect the cosmic
evolution at all, and hence does not affect the solution $H(t)$ of
the Friedman equation. Here we shall quantify this claim and prove
that this behavior is indeed what happens. Let us consider the
first Friedman equation (\ref{eqnsofmkotion}), for the $R^2$
$f(R)$ gravity (\ref{starobinsky}) and with the axion scalar field
potential satisfying the condition (\ref{conditiononaxion2}),
which holds true for all the cosmological eras during and after
inflation. In this case, the first Friedman equation of Eq.
(\ref{eqnsofmkotion}) becomes,
\begin{equation}\label{friedmanequationinflation}
3H^2\left(1+\frac{1}{18H_i^2}R+h(\phi)G'(R)\right)\simeq
\frac{1}{2}\kappa^2\dot{\phi}^2+\frac{1}{72H_i^2}R^2+2V\kappa^2-\frac{1}{6H_i^2}H\dot{R}-3H\dot{R}G''(R)h(\phi)\,
.
\end{equation}
During inflation, the curvature is of the order $R\sim 12H_I^2\sim
\mathcal{O}(1.2\times 10^{45})$eV, and thus the term containing
$\sim G'(R)$ in the left hand side of Eq.
(\ref{friedmanequationinflation}) is $\sim R^{\gamma-1}$ so it is
extremely suppressed during inflation. The same applies for the
term $\sim G''(R)$ in the right hand side, so these two terms can
be omitted. As for the term $\frac{1}{2}\kappa^2\dot{\phi}^2$, in
view of the initial conditions (\ref{axioninitialconditions}), by
taking $\zeta \sim \mathcal{O}(10^{-10})$ for example, this is of
the order $\frac{1}{2}\kappa^2\dot{\phi}^2 \sim
\mathcal{O}(8\times 10^{-74})$eV$^2$, and also the axion potential
term is of the order $2V\kappa^2\sim \mathcal{O}(3\times
10^{-38})$eV$^2$. In contrast, the $\sim R^2$ term is of the
order, $\frac{1}{72H_i^2}R^2\sim \mathcal{O}(2\times
10^{62})$eV$^2$, where we took $H_i\sim \mathcal{O}(10^{13})$eV
for phenomenological reasons \cite{Odintsov:2015gba}. Therefore,
comparing the scalar field and the $f(R)$ gravity terms during
inflation, it is obvious that the inflationary era is overwhelmed
by the Starobinsky model, so the Hubble rate is determined by the
$f(R)$ gravity. Thus the Friedman equation becomes approximately,
\begin{equation}\label{patsunappendixinflation}
\ddot{H}-\frac{\dot{H}^2}{2H}+3H_i^2H=-3H\dot{H}\, ,
\end{equation}
and by assuming a slow-roll evolution, the above equation becomes,
\begin{equation}\label{patsunappendix1inflation}
3H_i^2H=-3H\dot{H}\, ,
\end{equation}
which has the following quasi-de Sitter evolution as solution,
\begin{equation}\label{quasidesitter}
H(t)=H_0-H_i^2 t\, .
\end{equation}
The model thus results to the well-know spectral index of
primordial curvature perturbations $n_s$ and the tensor-to-scalar
ratio,
\begin{equation}\label{starobinskymodelobservationalindices}
n_s\sim 1-\frac{2}{N},\,\,\,r\sim \frac{12}{N^2}\, ,
\end{equation}
where $N$ is the $e$-foldings number. The phenomenology of the
Starobinsky model is the most successful among $f(R)$ gravity
models, so we do not discuss the inflationary era further. The
purpose of this section was to evince that the axion field does
not affect the cosmological evolution at all, during the
inflationary era, because it is frozen in its vacuum expectation
value, and its dynamics make it extremely overdamped, allowing the
$f(R)$ gravity to utterly control the the dynamical evolution of
the Universe. Thus the background geometry is controlled solely by
the $f(R)$ gravity during inflation.

\subsection{The Dark Energy Era}

In the previous subsection we showed that the $f(R)$ gravity will
dominate the Universe's evolution during the inflationary era, so
long as the axion field remains frozen around its vacuum
expectation value. However, as the Universe expands, when $H\sim
m_a$, the axion field will start oscillating, thus its dynamics
will change, since it is not frozen anymore, and following the
considerations of section III-A, its dynamical evolution is given
in Eq. (\ref{solutionaxionradandaft}) and also by assuming a
slow-varying evolution, the axion field energy density will behave
as $\rho_a\sim a^{-3}$. Effectively, this indicates that the axion
field for cosmic times corresponding to an era  for which $m_a\gg
H$, will behave as dark matter. The background evolution for the
eras between inflation and the dark energy era will be controlled
possibly by both the $f(R)$ gravity, the non-minimal coupling
$h(\phi)G(R)$ and the axion field, however our interest in this
subsection is on the late-time era, where the curvature is
significantly small. For clarity let us quote here the Friedman
equation and we discuss the significance of the various terms
appearing in it. The Friedman equation reads,
\begin{equation}\label{friedmanequationinflationdarkenergy}
3H^2\left(1+\frac{1}{18H_i^2}R+h(\phi)G'(R)\right)\simeq
\frac{1}{2}\kappa^2\dot{\phi}^2+\frac{1}{72H_i^2}R^2+2V\kappa^2-\frac{1}{6H_i^2}H\dot{R}-3H\dot{R}G''(R)h(\phi)\,
,
\end{equation}
and it is clear that the dominant term in the left hand side of
the above differential equation is $\sim h(\phi)G'(R)$ because it
contains the term $G'(R)\sim R^{\gamma-1}$ and since
$0<\gamma<0.75$ it contains negative powers of the curvature. On
the right hand side, the dominant term is solely
$-3H\dot{R}G''(R)h(\phi)$ for the reason that it contains
$G''(R)\sim R^{\gamma-2}$ which is strongly dominant for small
curvatures. The potential term $2V\kappa^2$ is proportional to
$\sim A^2$ and the kinetic term $\frac{1}{2}\kappa^2\dot{\phi}^2$
is depends on powers of $A$ and $\dot{A}$, and since $A\sim
a^{-3/2}$ these terms are subdominant to the term $\sim G''(R)$.
The same applies to the term $\sim R^2$. To have a strong idea on
how large is the term $\sim R^{\gamma-2}$ recall that the Hubble
rate today is approximately $H_0\simeq 10^{-33}$eV so by choosing
for example $\gamma =0.74$, the term $G''(R)$ becomes $G''(R)\sim
R^{\gamma-2}\sim \mathcal{O}(6.3\times
10^{81})$$\left(\mathrm{eV}\right)^{4-2\gamma}$, while the $R^2$
term is of the order $\frac{1}{72H_i^2}R^2\sim \mathcal{O}(2\times
10^{-178})$eV$^2$, which is extremely small. Thus, the Friedman
equation at late-times becomes,
\begin{equation}\label{friedmanequationinflationdarkenergy}
3H^2h(\phi)G'(R)\simeq -3H\dot{R}G''(R)h(\phi)\, ,
\end{equation}
which by using $G(R)\sim R^{\gamma}$, can be written as follows,
\begin{equation}\label{finalfriedmanequationlatetimes}
RH\simeq \left(1-\gamma \right)\dot{R}\, ,
\end{equation}
so by using $R=12H^2+6\dot{H}$ for the FRW Universe and by
omitting the term  $\sim H^3$ which is subdominant at late times,
the solution is,
\begin{equation}\label{finalhubblerate}
H(t)\simeq \frac{\sqrt{2} \sqrt{1-\gamma } \sqrt{\Lambda
}}{\sqrt{3-4 \gamma }}\tanh \left(\frac{1}{2} \left(\frac{\sqrt{2}
\sqrt{3-4 \gamma } \Theta  \sqrt{\Lambda }}{\sqrt{1-\gamma
}}+\frac{\sqrt{2} \sqrt{3-4 \gamma } \sqrt{\Lambda }
t}{\sqrt{1-\gamma }}\right)\right)\, ,
\end{equation}
where $\Lambda$ and $\Theta$ are integration constants. Also from
the above solution, the parameter $\gamma$ is constrained to take
values in the interval $0<\gamma<0.75$. Effectively, at late times
which correspond to large cosmic times, say at $t=t_0$, the Hubble
rate is approximately constant,
\begin{equation}\label{latetimedesitterevolutin}
H(t_0)\simeq \frac{\sqrt{2} \sqrt{1-\gamma } \sqrt{\Lambda
}}{\sqrt{3-4 \gamma }}\, .
\end{equation}
Therefore the late-time evolution is a de Sitter evolution, that
the Universe evolves in an accelerating way. This can also be seen
by calculating the deceleration parameter
$q=-1-\frac{\dot{H}}{H^2}$, which for the Hubble rate
(\ref{finalhubblerate}) is equal to,
\begin{equation}\label{decelerationparameter}
q=-1-\frac{(3-4 \gamma ) \mathrm{csch}^2\left(\frac{1}{2}
\left(\frac{\sqrt{2} \sqrt{3-4 \gamma } \Theta  \sqrt{\Lambda
}}{\sqrt{1-\gamma }}+\frac{\sqrt{2} \sqrt{3-4 \gamma }
\sqrt{\Lambda } t}{\sqrt{1-\gamma }}\right)\right)}{2 (1-\gamma
)}\, ,
\end{equation}
which is clearly negative during the late-time era, thus an
accelerating evolution occurs. Therefore, the axion coupling
$h(\phi)G(R)$ affects the late-time era in a dominant way, since
it controls the evolution making the Universe to accelerate. In
effect, in some sense the microphysics primordial coupling
$h(\phi)G(R)$ which was subdominant at the high curvature era and
during inflation, becomes dominant at late times, causing the
Universe to accelerate and thus providing a dark energy era. Also
the effective equation of state of the Universe at late-times is,
\begin{equation}\label{eosfinaltimes}
w_{eff}=-1-\frac{(3-4 \gamma ) \text{csch}^2\left(\frac{1}{2}
\left(\frac{\sqrt{2} \sqrt{3-4 \gamma } \Theta  \sqrt{\Lambda
}}{\sqrt{1-\gamma }}+\frac{\sqrt{2} \sqrt{3-4 \gamma }
\sqrt{\Lambda } t}{\sqrt{1-\gamma }}\right)\right)}{3 (1-\gamma
)}\, ,
\end{equation}
so this becomes approximately $w_{eff}\sim -1$, due to the
exponential decay of the hyperbolic cosecant function for large
cosmic times.

Let us here quote an interesting phenomenological observation. By
taking into account that today $H_0\sim 10^{-33}$eV, for
$\gamma=0.74$ we get the approximate value of the integration
constant $\Lambda=7.69231\times 10^{-68}$eV$^2$. Also for
$\gamma=0.2$ we have $\Lambda=1.375\times 10^{-66}$eV$^2$ and it
can be shown that for $\gamma$ belonging in the interval
$0<\gamma<0.75$, so when $0<\gamma \leq 0.74$, the parameter
$\Lambda$ takes values in the interval $1.5\times
10^{-66}\mathrm{eV}^2<\Lambda < 7.69231\times 10^{-68}$eV$^2$. The
value of the actual cosmological constant today is approximately
$\Lambda_0\sim 10^{-66}$eV$^2$, so it is quite intriguing that for
the allowed values of the dimensionless parameter $\gamma$, the
integration constant $\Lambda$ takes values quite close to the
actual value of the cosmological constant $\Lambda_0$ today, and
also it has the same dimensions eV$^2$. In addition, as the
parameter $\gamma$ approaches the value $0.75$, then the constant
parameter $\Lambda$ takes smaller values, for example, as we also
noted in the introduction, when $\gamma=0.74999999999$, then
$\Lambda=8\times 10^{-77}$eV$^2$. So the question is, is there any
possibility that the relation between the current Hubble rate and
the cosmological constant is given by Eq.
(\ref{latetimedesitterevolutin})? It is certainly known that for
the present day cosmological constant it holds true that
$\Lambda_0\sim H_0^2$ so this is an intriguing result. However, as
$\gamma$ approaches the limiting value $\gamma=0.75$, the
parameter $\Lambda$ takes much more smaller values in comparison
to the cosmological constant today.

Before closing, let us note that a similar phenomenology can be
obtained by using a pure $f(R)$ gravity model in addition to the
axion scalar field, without the axion non-minimal coupling. In
this case the gravitational action would be of the form,
\begin{equation}
\label{mainactiongravitationalnew} \mathcal{S}=\int
d^4x\sqrt{-g}\left[
\frac{1}{2\kappa^2}\left(R+\frac{1}{36H_i^2}R^2-\delta R^{\gamma}
\right)-\frac{1}{2}\partial^{\mu}\phi\partial_{\mu}\phi-V(\phi)
\right]\, .
\end{equation}
It can also be shown in this case that the $R^2$ gravity dominates
inflation, and also at late times, the term $R^{\gamma}$ dominates
the evolution. However as we show in the next section, the
non-minimally coupled axion model predicts also a stiff matter era
preceding the inflationary era, which is not predicted from the
model (\ref{mainactiongravitationalnew}).

Before closing, some interesting questions must be discussed.
Firstly, the models (\ref{mainaction}) and
(\ref{mainactiongravitationalnew}) produce qualitatively similar
results for the dark energy era, and specifically for the
late-time de Sitter solution, but we did not quote these for
brevity. However, for the model (\ref{mainactiongravitationalnew})
the pre-inflationary stiff era for the axion would absent.
Secondly and important question is what are the predictions of the
non-minimally coupled model (\ref{mainaction}) at a perturbative
level, since all the study was performed at a background level, so
what would be the effects of the non-minimal coupling at a
perturbative level, and in effect, what would be the effects of
this axion dark matter coupling to dark energy for structure
formation? This question is quite interesting, though non-trivial
to address directly in this paper. The perturbations for the
combined theory for eras corresponding to cosmic times after
inflation and before the dark energy era, mainly correspond to an
$f(\phi,R)$ theory and were calculated in detail in
\cite{Hwang:2005hb}, so in principle one can have a concrete
picture on how the perturbations at intermediate eras evolve
exactly. However, since during inflation the $R^2$ gravity
dominates, the perturbative modes during the horizon exit, evolve
in a similar way as in the pure $R^2$ model. An exact calculation
in the context of $f(\phi,R)$ performed in \cite{Hwang:2005hb},
reveal the exact behavior of the perturbations evolution.
Moreover, after inflation, the reheating picture changes
drastically, so one must know the exact behavior of the
$f(\phi,R)$ theory, in order to correctly study the reheating era
and the subsequent matter formation era, this is not easy trivial
to do analytically though. There is also another feature to be
pointed out. The axion may be overdamped during inflation, but it
actually acquires perturbations during inflation, so it may
directly affect the primordial curvature perturbation modes, and
hence affect their evolution, after horizon exit, as we already
mentioned above. Although the effect are moderate, in an exact
theory, these should be take into account. The theory we studied
though is an effective theory, so we did not take these effects
into account. Also after inflation, the axion energy density
behaves as $\rho_a\sim a^{-3}$, but this is an average behavior
for the axion field, and the average value may vary randomly
within the observable Universe, as the assumed location of the
observable Universe is altered in a comoving volume. Thus, locally
the axion physics may alter, when for example intergalactic scales
are considered. The physics of this discussion is highly
non-trivial to even discuss in the context of the effective model
we discussed, however these are noteworthy questions that should
be addressed appropriately in future works.

\subsection{A Prediction of the Theory: Stiff Pre-inflationary Era}

Up to now we assumed that the condition of Eq.
(\ref{conditionforaxion1}) or equivalently Eq.
(\ref{conditiononaxion2}) was holding true for cosmic times during
and after the inflationary era. It is though conceivable that at
some cosmic time before inflation, the potential could take
smaller values in comparison to the value it takes when the axion
field takes its vacuum expectation value corresponding to the
broken Peccei-Quinn symmetry. In effect, at some pre-inflationary
primordial era we could have $2V(\phi)\sim
\frac{1}{\kappa^2}h(\phi)G(R)$. This can be indeed true if the
$U(1)$ breaking occurs during a second order phase transition in
the axion field, so that the potential continuously deforms to the
value $\sim \frac{1}{2}m_a^2\phi_i^2$. When the condition
$2V(\phi)\sim \frac{1}{\kappa^2}h(\phi)G(R)$ holds true, the
differential equation (\ref{scalarfieldeqn}) that governs the
evolution of the scalar field becomes,
\begin{equation}\label{scalarfieldeqnstiff}
\ddot{\phi}+3H\dot{\phi}=0\, ,
\end{equation}
so by solving this we get,
\begin{equation}\label{solstiff}
\dot{\phi}\sim a^{-3}\, ,
\end{equation}
where $a$ is the scale factor, and in effect we have
$\dot{\phi}^2\sim a^{-6}$. Due to the condition $2V(\phi)\sim
\frac{1}{\kappa^2}h(\phi)G(R)$, the energy density of the axion
scalar field at this primordial era is approximately $\rho_a\sim
\dot{\phi}^2$, so in view of the solution (\ref{solstiff}) we
found, we have $\rho_a\sim a^{-6}$. This behavior is
characteristic of a stiff matter fluid and thus the model predicts
a stiff matter era before the inflationary era. It is intriguing
to note that Zel'dovich had introduced the hypothesis that a stiff
matter era occurred in the primordial Universe
\cite{Zeldovich:1972zz}, however in the Zel'dovich the matter
fluid consisted of a stiff cold baryon gas, whereas in our case
the stiff matter fluid is the axion itself. Hence, the presence of
the non-minimal coupling apart from affecting the late-time era,
also affects the pre-inflationary primordial era, allowing for a
stiff matter era to be realized.

\section{Concluding Remarks}

The possibility of detecting the axion field in the next decades
apart from exciting, is perhaps the last resort of dark matter
particle physics. This is due to the fact that the experiments
searching for large mass weakly interacting massive particles seem
not to produce any hint for the existence of a dark matter
particle, so unless some evidence for supersymmetry is found, one
has a last possibility for detecting dark matter particles having
extremely low masses. The misalignment axion model seems to be an
appealing candidate, and in this paper we provided an effective
theory that may unify the inflationary era with the dark energy
era, and finally describe a dark matter dominated era.
Particularly, we assumed that the $f(R)$ gravity is described by
the $R^2$ model at early times and also contains a non-minimal
coupling to the axion scalar field of the form
$h(\phi)R^{\gamma}$. Due to the fact that the axion field remains
frozen in its primordial vacuum expectation value, the Starobinsky
model controls the early-time dynamics, providing a viable
quasi-de Sitter evolution. As the Universe expands, the axion
field starts to oscillate, approximately when $H\sim m_a$. By
assuming a slowly-varying evolution, the axion field energy
density scales as $\rho_a\sim a^{-3}$, which describes a dark
matter effective fluid, regardless what the actual Hubble rate of
the background is. At late times, the non-minimal coupling
dominates the evolution, and the resulting Hubble rate at late
times is nearly a de Sitter one, thus describing an accelerating
Universe. Also the theory predicts the existence of a stiff matter
era for the axion field at a primordial pre-inflationary era, in
which case the energy density scales as $\rho_a\sim a^{-6}$.

It is also known that in the context of the singular inflation
scenario, one can unify the early-time acceleration with late-time
acceleration in $f(R)$ gravity \cite{Odintsov:2016plw}. Hence, it
is natural to discuss the unification of singular-inflation with
dark energy, including above the axion dark matter and non-minimal
coupling terms. In the same way one can take into account
logarithmic quantum gravity corrections in order to unify
inflation with dark energy, as it was done without the axion dark
matter field in Ref. \cite{Odintsov:2017hbk}. The inclusion of the
axion dark matter field may strengthen the viability of these
theories, providing a dark matter era and a smooth transition from
the dark matter era to the dark energy era.

An interesting issue which we did not discuss, is the radiation
domination era and specifically the reheating issue in the dawn of
this era. Particularly, the axion oscillations may amplify the
curvature oscillations caused by the $R^2$ gravity, thus altering
the reheating temperature. Work is in progress towards this
research line.

\section*{Acknowledgments}

This work is supported by MINECO (Spain), FIS2016-76363-P, by
project 2017 SGR247 (AGAUR, Catalonia) (S.D.O) and by Russian
Ministry of Science and High Education, project No. 3.1386.2017.

\end{document}